\documentclass[twocolumn]{bmcart}

\usepackage{amsthm,amsmath}
\usepackage{amsfonts}
\usepackage{commath}

\usepackage{booktabs}

\usepackage{multirow}
\RequirePackage[authoryear]{natbib}

\usepackage[utf8]{inputenc} 
\usepackage{microtype}
\usepackage{tensor}
\usepackage{graphicx}

\startlocaldefs
\newcommand\tens[2][2]{
\ifthenelse{#1=4} 
{\ensuremath{\mathbb{#2}}}
{\ensuremath{\boldsymbol{#2}}}}

\DeclareMathOperator{\grad}{\! \nabla}

\newcommand{\T}{^{\sf T}}
\DeclareMathOperator*{\argmin}{arg\,min}

\newcommand{\mrm}[1]{\mathrm{#1}}
\newcommand{\lb}{\left(}
\newcommand{\rb}{\right)}

\newcommand{\fe}{\tens{e}}

\newcommand{\cP}{{\mathcal P}}

\newcommand{\mepsR}{\tens{\varepsilon}{^{\mrm R}}}
\newcommand{\mgomega}{\nabla \tens{\omega}}

\newcommand{\fsigma}{\tens{\sigma}}
\newcommand{\feps}{\tens{\varepsilon}}
\newcommand{\fS}{\tens{S}}
\newcommand{\fmu}{\tens{\mu}}

\newcommand{\Gc}{G_\mrm{c}}

\endlocaldefs

\begin{document}
\begin{frontmatter}

\begin{fmbox}
\dochead{Research}


\title{Determining Cosserat constants of 2D cellular solids from beam models}


\author[
   addressref={aff1},                   
   corref={aff1},                       
   email={stefan.liebenstein@fau.de}   
]{\inits{SL}\fnm{Stefan} \snm{Liebenstein}}
\author[
   addressref={aff1,aff2},
]{\inits{MZ}\fnm{Michael} \snm{Zaiser}}


\address[id=aff1]{
  \orgname{Institute of Materials Simulation (WW8), Friedrich-Alexander University Erlangen-N\"urnberg (FAU)}, 
  \street{Dr.-Mack-Strasse 77},                     %
  \postcode{90762}                                
  \city{F\"urth},                              
  \cny{Germany}                                    
}
\address[id=aff2]{
  \orgname{Department of Engineering and Mechanics, Southwest Jiaotong University}, 
  \city{Chengdu}
  \cny{P.R. China}                                    
}





\begin{abstractbox}

\begin{abstract} 
We present results of a two-scale model of disordered cellular materials where we describe the microstructure in an idealized 
manner using a beam network model and then make a transition to a Cosserat-type continuum model describing the same material on
the macroscopic scale. In such scale transitions, normally either bottom-up homogenization approaches or top-down reverse 
modelling strategies are used in order to match the macro-scale Cosserat continuum to the micro-scale beam network. Here we use
a different approach that is based on an energetically consistent continuization scheme that uses data from the beam network model 
in order to determine continuous stress and strain variables in a set of control volumes defined on the scale of the individual 
microstructure elements (cells) in such a manner that they form a continuous tessellation of the material domain. Stresses and 
strains are determined independently in all control volumes, and constitutive parameters are obtained from the ensemble of control 
volume data using a least-square error criterion. We show that this approach yields material parameters that are for regular honeycomb structures in close agreement with analytical results. For strongly disordered cellular structures, the thus parametrized Cosserat 
continuum produces results that reproduce the behavior of the micro-scale beam models both in view of the observed strain patterns and 
in view of the macroscopic response, including its size dependence. 

\end{abstract}


\begin{keyword}
\kwd{cellular materials}
\kwd{disorder}
\kwd{micro-to-macro transition}
\kwd{Cosserat continuum}
\end{keyword}


\end{abstractbox}
\end{fmbox}

\end{frontmatter}



\section{Introduction}
Classical constitutive models for elastic materials behavior are local and do not possess intrinsic length scales. For cellular solids the inadequacy of such models is obvious once the relevant scale of a problem becomes comparable to the intrinsic length scale defined by the cell size or strut lengths can be about the same order of magnitude as the system size. As a result smaller samples may behave differently compared to larger ones.

Such size effects can be captured by material models which account for internal length scales, such as constitutive equations in higher order continuum theories \citep{altenbach2013}.  The idea goes back to \citet{cosserat1909} who enriched the classical formulation by additional independent rotational degrees of freedoms (micro-rotations) and their gradients. Later this was generalized  by \citet{eringen1964, eringen1966} to arbitrary, independent micro-deformations. The extended modelling capabilities of such generalized continuum theories come with a cost: the generalized theories require additional boundary conditions and the extended constitutive laws may, in the most general cases, involve very significant numbers of material parameters. Experimental determination of these parameters is much more difficult and error-prone compared to the classical theory, efforts have therefore focused on the Cosserat continuum as the simplest case. We mention attempts by \citet{gauthier1975} for torsion of a circular cylinder as well as a series of studies by Lakes and co-workers \citep{yang1982, lakes1983, lakes1986, chen1991, anderson1994} who investigated cellular materials including bone, polymeric and metallic foams with regards to size effects.  More recently Wheel and co-workers \citep{McGregor2014, beveridge2013, Wheel2015} combined experiments and finite element simulations to establish constitutive parameters for Cosserat-type models of heterogeneous materials.

Multi-scale simulation methods which relate a micro-scale model with full resolution of the material microstructure to a generalized continuum macro-model may serve as an alternative or additional approach to experimental parameter identification. A top-down multi-scale approach consists in utilizing results from the micro-scale model in a similar manner as experimental data to fit parameters of the generalized continuum for a matching overall system response \citep{tekoglu2008, diebels2002, mora2007}. Alternatively, bottom-up homogenization methods can be used to transfer informations from the underlying microstructure into the continuum model. Different approaches exists, notably asymptotic methods \citep{Ghosh1996, forest2001, dos2012}, polynomial expansion \citep{forest1998, forest1998b} and numerical Cauchy-continuum to higher order continuum homogenizations \citep{Feyel2003, kouznetsova2004, Jaenicke2009}. Many of these approaches are limited to  regular or periodic lattices or are computationally expensive and thus not well suited for studies of stochastic systems, where many different microstructures are needed for statistically meaningful averages. The problem is particularly pronounced in  analysis of strongly disordered materials where the local response may be controlled by heterogeneous meso-scale stress and strain patterns, which pose inherent problems to homogenization schemes that rely on a certain regularity and macro-homogeneity of the material response. 

In this work we focus on two-scale modelling of cellular microstructures of variable disorder where on the microstructure level we represent the materials by networks of Timoshenko beams and on the macroscale by a Cosserat continuum. We first briefly recap the method we use to create stochastic microstructures. We then introduce a microstructure dependent continuization approach we have discussed in more detail elsewhere \citep{liebenstein2017}. This semi-analytical, energy based ansatz is used to obtain, from local stresses and deformations in the discrete model, continuous fields for a micropolar description of the deformation state in terms of displacement gradients, micro- and macro-rotations, and micro-rotation gradients, and of the stress state in terms of the corresponding conjugate stress variables. Linear least square inspired minimization is used for fitting representative, macroscopic material parameters to these data. Our investigations cover the spectrum from regular honeycombs to random cellular structures. The resulting constitutive parameters are validated by comparing global spatial stress and strain patterns to those derived from a direct implementation of a Cosserat continuum for compressive and simple shear loading of finite samples.

\section{Model}
\subsection{Computational generation of the microstructure}
Most cellular solids do not possess regular, periodic micro-structures such as in honeycombs or other lattice-like structures, but rather  exhibit stochastic cellular patterns with varying degree of statistical irregularity. To model the behavior of such stochastic microstructures in a statistically meaningful manner one has to rely on ensembles of statistically equivalent samples, rather than on single realizations.  To generate such ensembles we use the approach of \citet{gibson1999, zhu2000, tekoglu2005} to represent the cellular microstructure as a two-dimensional network of Timoshenko beams for which the network structure is generated via Voronoi tessellation of the system plane. The Voronoi tessellation represents reasonably the foaming process under the assumptions \citep{boots1982, vanderburg1997} that
\begin{enumerate}
\item all nuclei appear simultaneously,
\item the nuclei remain fixed in position throughout the grow process,
\item each nucleus grows isotropically, i.e., at the same rate in all directions,
\item the growth stops for each cell whenever it comes into contact with a neighbouring cell.
\end{enumerate}
To tune the degree of irregularity we adopt a method originally proposed by \citet{vanderburg1997}. As a starting point seeds are generated on a triangular lattice with lattice constant $\Delta p$, which results in a Voronoi tessellation consisting of regular honeycombs. Irregular systems are created by perturbing the position $\tens{p}$ of each seed by a stochastic vector $\delta  \tens{p}$. In order to obtain an spatially isotropic distribution the direction of the perturbation is chosen isotropically, whereas the perturbation distance $|\delta \tens{p}|$ is assumed to be exponentially distributed:
\begin{align}
f(\frac{|\delta \tens{p}|}{\Delta p},\beta) =  \frac{1}{\beta} \exp \lb -\frac{1}{\beta}\frac{|\delta \tens{p}|}{\Delta p} \rb.
\end{align}
The disorder parameter $\beta>0$ defines both the mean value and the standard deviation of the distribution. 

Size effects are one main characteristic of materials and models with an intrinsic length scale. In order to faithfully compare different system sizes one must ensure that all systems have the same relative density $\rho_\mrm{rel}$ and the same aspect ratio $R=W/H$ between the system width $W$ and the system height $H$. Under the assumption that all $N_\mrm{B}$ beams have the same in-plane width $w$ and out of plane thickness $t$ but vary in length $l_i \gg w$, the total volume covered by the beams is
\begin{align}
V_\mrm{B} = t w \sum_{i=1}^{N_\mrm{B}} l_i.
\end{align}
The relative density of a system is
\begin{align}
\rho_\mrm{rel} = \frac{V_\mrm{B}}{t W H} = \frac{w}{W H} \sum_{i=1}^{N_\mrm{B}} l_i,
\end{align}
which is kept constant at $\rho_\mrm{rel}=0.1$ by proportional scaling of $W$ and $H$.
As a default aspect ratio for the systems we use the aspect ratio  $R_0 = 2/\sqrt{3}$ of a single regular honeycomb. Each Timoshenko beam is assumed to be linear elastic with Young's modulus $E_{\mrm B}= 0.1$ GPa and  Poisson's ratio $\nu_{\mrm B}=0.3$. The cross-section is quadratic: $t=w=0.05 \Delta p$ with a shear coefficient proposed by \cite{cowper1966} of $\kappa =10+10 \nu_{\mrm B}/(12 + 11 \nu_{\mrm B}) \approx 0.85. $

The surface response of regular honeycombs strongly depends on the way boundaries are located with respect to the honeycomb lattice, i.e., on the length and orientation of beams intersecting the boundary. Consider a system where the end points of the beams are located exactly on the boundary, which corresponds to closed honeycombs at the boundary. Small perturbations of this configuration may open up the honeycomb structure which results in the unphysical effect that, in small systems where surface effects are non negligible, such small perturbations may lead to significantly different system stiffnesses. In regular honeycombs such effects may be desirable and even engineered, however, if we want to compare regular with random cellular structures we need to average them out. This can be done by averaging over multiple realizations where we shift the system boundaries by random fractions of the honeycomb lattice period,
\begin{align}
&\Delta y = 0.5 \Delta p  \,R_{\mrm h}(-1,1),\nonumber\\
&\Delta x = (0.5+\frac{ 1}{\sqrt{3}} )\Delta p \, R_{\mrm w}(-1,1),
\end{align}
where the random numbers $R_{\mrm w}(-1,1), R_{\mrm h}(-1,1)$ are uniformly distributed between $-1$ and $1$. Such a realization can be envisaged as a randomly located cut-out from a larger system. We therefore speak of random cut-outs when referring to the randomly chosen location of system boundaries with respect to the lattice underlying our beam network construction.

In this work we study two loading cases: displacement driven compression and shear. For a homogeneous continuum, these loadings would correspond to pure uniaxial compression and to simple shear loading. The displacements $\tens{u}$ and the rotations $\tens{\phi}$ of the beams at the boundaries are given  in Table~\ref{tb:Dirichlet BC} and chosen such that they are comparable to experiments  \citep{andrews2001, tekoglu2011}.

\begin{table*}[h]
\begin{center}
\begin{tabular}{ l l l l l l}
\toprule
 Loading & Bottom ($y=0$) & Top ($y=H$) & Left ($x=0$) & Right ($x=w$) \\
  \midrule
  Compression & $\bar{u}_y=0$ & $\bar{u}_y= H / 0.05$ & free & free \\
  Simple Shear  & $ \bar{u}_x=0, \phi=0$  & $ \bar{u}_x= H/ 0.05 ,\phi=0$ & free& free\\
  \bottomrule\\
\end{tabular}
  \caption{Boundary conditions for the two  investigated loading conditions.}
  \label{tb:Dirichlet BC}
	\end{center}
\end{table*}

It should be noted that from a macroscopic point of view, homogeneous uniaxial compression of an isotropic continuum is not expected to produce higher order/size effects, because uniaxial loading does not induce micro-rotations \citep{kirchner2007}. On the meso-scale, however, local rotations are always present, and in strongly disordered systems they even might play an important role as a result of structural inhomogeneities. We therefore study both loading cases. 

\subsection{Continuum representaton of stresses and strains in the beam network}
The modelling and finite element simulation of cellular solids as a network of Timoshenko beams gives naturally forces and displacements, evaluated at nodes of the beam elements. To make the transition from this representation at discrete nodal points to a spatially continuous Cosserat continuum we use a method introduced by the authors \citep{liebenstein2017} that is based on the energy equivalence of the beam network and the contiuum. The balance equations for the Cosserat continuum are (see e.g. \citet{eringen1999, forest2009})
\begin{align}
\label{eq:linear_momentum}
\nabla .{\fsigma+\fS} &= \tens{0},\\
\label{eq:angular_momentum}
\nabla .{\fmu} - \fS: \tens{\epsilon}&= \tens{0},
\end{align}
where $\fsigma$ is the symmetric (Cauchy) stress and $\fS$ the skew-symmetric stress tensor which balances the couple stress $\fmu$. The symbol  $\tens{\epsilon}$ denotes the Levi-Civita or permutation tensor, $\nabla$ the gradient operator, $.$ is the single contraction and $:$ the double contraction of two tensors. The corresponding stress traction and couple stress traction read
\begin{align}
\tens{t} &= \lb \fsigma+\fS \rb \T . \tens{n}, \\
\tens{m} &= \fmu \T . \tens{n}.
\end{align}
With the displacements $\tens{u}$ and the rotations $\tens{\omega}$ the corresponding work conjugated strains are related via the strain energy density $W_c$ 
\begin{align}
\feps &= \rho \dpd{W_c}{\fsigma} = \frac{1}{2} \left( \grad{\tens{u}} +  \grad{\tens{u}} \T \right), \label{eq:sym_strain} \\
\mepsR &= \rho \dpd{W_c}{\fS} = \frac{1}{2} \left( \grad{\tens{u}} -  \grad{\tens{u}} \T \right) + \tens{\epsilon}. \tens{\omega},\label{eq:skw_strain}  \\
\mgomega &= \rho \dpd{W_c}{\fmu}.\label{eq:rot_grad} 
\end{align}
Under the assumption that the virtual work $\delta W_\mrm{b}$ of the sum of all beams $N_\mrm{b}$ is the same as the virtual work of the continuum $\delta W_\mrm{c}$ in any given control volume  $V_\mrm{c}$, the following relations for the control volume averages can be derived \citep{liebenstein2017}:
\begin{align}
\langle \fsigma \rangle_\mrm{c} &=  \frac{1}{V_{\mrm c}} \int_{V_{\mrm c}} \fsigma \dif V = \mathrm{sym}\lb\frac{1}{V_{\mrm c}} \sum_{k=1}^{N_\mrm{b}}\tens{F}^k \otimes  \tens{l}^k\rb, \\
\langle \fS \rangle_\mrm{c} &=  \frac{1}{V_{\mrm c}} \int_{V_{\mrm c}} \fS \dif V= \mathrm{skw}\lb\frac{1}{V_{\mrm c}} \sum_{k=1}^{N_\mrm{b}}\tens{F}^k \otimes  \tens{l}^k\rb, \\
\langle \fmu \rangle_\mrm{c}
 &= \frac{1}{V_{\mrm c}} \int_{V_{\mrm c}} \fmu \dif V = \frac{1}{V_{\mrm c}} \sum_{k=1}^{N_\mathrm{b}}\lb \tens{M}^k 
- \tens{l}^k \times  \tens{F}^\mrm{k}\rb \otimes  \tens{l}^k, \label{eq:couple_stress}
\end{align}
where $\times$ denotes the cross-product, $\tens{F}$ is the beam force, $\tens{M}$ the moment acting on a triple (or higher) junction point and $\tens{l}$ is the so-called beam vector, which is the difference between the midpoint between two junctions and a junction point. The averaged displacement and rotation gradients are for an $N_c$-sided control volume calculated from the nodal displacements and rotations as
\begin{align}
\langle \tens{e} \rangle_{\mrm{c}} &= \frac{1}{V_\mrm{c}} \int_{V_\mrm{c}} \nabla \tens{u} \dif V\nonumber\\
&=  
\frac{1}{V_\mrm{c}} \sum_{k=1}^{N_\mrm{c}} \left(\frac{\tens{u}^{k} + \tens{u}^{k+1}}{2} \otimes \tens{n}^{(k,k+1)}\right) \tens{b}^{(k,k+1)},\\
\langle \mgomega \rangle_{\mrm{c}} &= \frac{1}{V_\mrm{c}} \int_{V_\mrm{c}} \mgomega  \dif V     \nonumber\\
&=  \frac{1}{V_\mrm{c}} \sum_{k=1}^{N_\mrm{c}} \left(\frac{\tens{\omega}^{k} + \tens{\omega}^{k+1}}{2} \otimes \tens{n}^{(k,k+1)}\right) \tens{b}^{(k,k+1)},
\label{eq:strainavs}
\end{align}
where $\tens{u}^k$ is the displacement and $\tens{\omega}^{k}$  the rotation of corner node number $k$ in a clockwise enumeration, $b^{(k,k+1)}$ is the length of the polygon side which connects nodes $k$ and $k + 1$, and $\tens{n}^{(k,k+1)}$ is the outward pointing normal vector to this side. From the displacement gradients, strain-like quantities derive in analogy to equations \eqref{eq:sym_strain}, \eqref{eq:skw_strain} as
\begin{align}
\langle \feps \rangle_\mrm{c} &= \frac{1}{2} \lb \langle \tens{e}  \rangle_{\mrm{c}} + \langle \tens{e}  \rangle_{\mrm{c}}\T \rb, \\
\langle \mepsR \rangle_\mrm{c} &= \frac{1}{2} \lb \langle \tens{e} \rangle_{\mrm{c}} - \langle \tens{e}  \rangle_{\mrm{c}}\T \rb + \tens{\epsilon}. \langle \tens{\omega} \rangle_\mrm{c}.
\end{align} 
The averaged control volume rotation $\langle \tens{\omega} \rangle_{\mrm{c}}$ is calculated by a linear interpolation of the beam-midpoint rotations to be consistent with the constant rotation gradient. In addition to the control volume averages, the system average of a quantity $(\cdot)$ can be approximated as
\begin{align}
\langle \cdot \rangle_\mrm{S} \approx  \frac{1}{V_{\mrm S}} \sum_{k=1}^{N_{\mrm v}} V_{\mrm c}^k \langle \cdot \rangle_\mrm{c}^k,
\label{eq:system_avg}
\end{align}
where $V_{\mrm S} = \sum_{k=1}^{N_{\mrm v}} V_c^k$ is the system volume and $ N_{\mrm v}$ the number of control volumes in the system.

The control volumes are constructed to match the local microstructure. Each control volume is associated to a triple (or higher) junction in the beam network. The control volume corners are the midpoints between the chosen junction point and the connected neighbour junctions. Additional corner points located at the center points of the Voronoi tessellation ensure that the control volumes provide a tessellation of the entire domain. The exact details of the construction process are given by  \citet{liebenstein2017}.
\begin{figure*}
\centering
\includegraphics[width=0.9\textwidth]{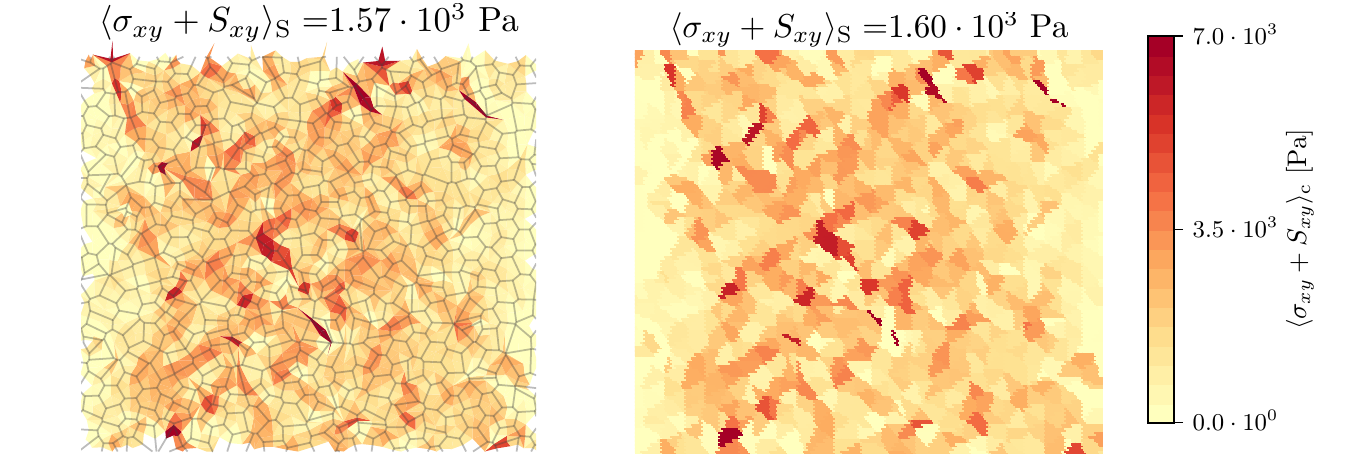}
\caption{Shear stresses in the control volumes (left) and nearest-neighbour interpolation (right) for a beam network under shear loading. In the uninterpolated control volume representation the micro-structure is shown in the background in grey. System parameters: $H=19 \Delta p$ and $\beta=0.3$.}
\label{fig:interpolation}
\end{figure*}

\subsection{Identifying Cosserat parameters}
Because of the randomness of the underlying micro-structure each realization is locally inhomogeneous and also anisotropic regarding its local material properties. A macroscopic representative system however exhibits homogeneous and isotropic material behavior. 
This means that we can not compare a single realization with a higher order continuum but rather we need to average over an ensemble of many realizations: The ensemble average restores the statistical homogeneity and isotropy of material behavior. Because local micro-structural features are less important in bigger systems, for such systems less realizations are needed to evaluate ensemble averages in a statistically reliable manner. In the following we average over 200 different realizations for $H \leq 25 \Delta p$ and over 50 realizations for $H > 25 \Delta p$. For all irregular systems 20 random cut-outs are chosen for each realization. Because regular systems where $\beta=0$ have a uniform microstructure we increase the number of random cut-outs to 200 when performing averages for such systems. 
 
\begin{figure*}
\centering
\includegraphics[width=0.8\textwidth]{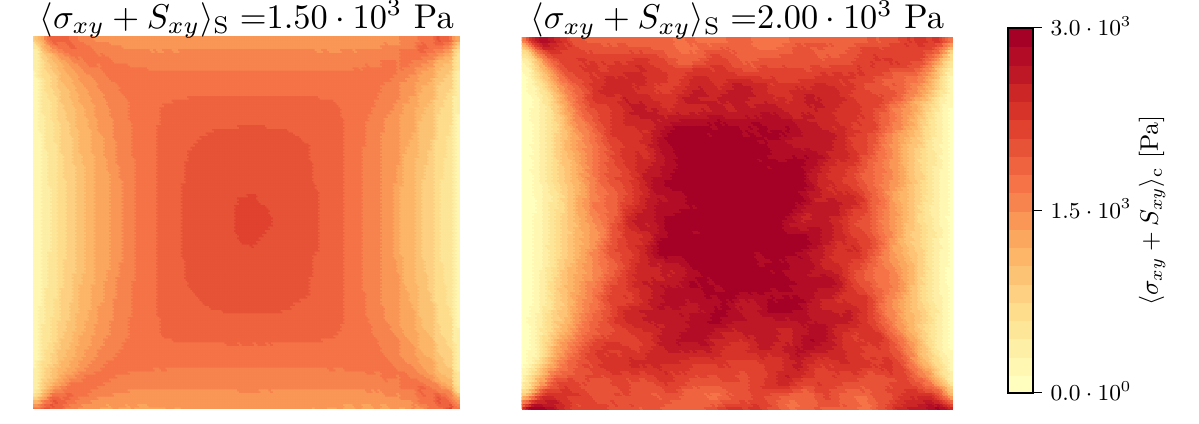}
\caption{Averaged system response of the beam network for regular $\beta=0$ and irregular $\beta=5$ microstructures, simple shear loading by rigid displacement of the top surface by $0.05H$, system height $H=19$; note the more rigid response of the disordered system.}
\label{fig:averaged_system}
\end{figure*}

For the averaging procedure we first notice that by construction the quantities are piece-wise constant in the control volumes which are system specific. To perform averages we need to interpolate them to a common grid. In order to retain the local structure of the stress and strain fields of each realisation we use a nearest neighbor interpolation to a fine grid such that there are approximately 4 interpolation grid points per control volume. This ensures that the localized structure of the stress and strain fields of each realization is preserved as shown in \figref{fig:interpolation}. The resulting interpolations use a common grid and can be directly averaged to analyze the average system response. Examples of the resulting, ensemble averaged stress field for regular and irregular systems are depicted in \figref{fig:averaged_system}. 

The averaged system has its counterpart in a higher order continuum for which we want to determine effective material parameters. While the effective classical material parameters can be identified via direct analysis of reaction forces and boundary displacements (similar to  Eq. \eqref{eq:reaction_force_beam}) the Cosserat material constants are not as easily accessible. However, our continuization method computes stress-like $\{\fsigma, \fS, \fmu \}$ and strain-like quantities $\{\feps, \mepsR, \mgomega \}$ independently and thus allows for an identification of material properties. As showed by \citet{warren1987,gibson1999}, periodic regular honeycombs are (macroscopically) isotropic. Completely stochastic systems of Voronoi cells are expected to be isotropic as well, similar to multi-grain materials, where each grain itself is anisotropic, but on average the system behaves isotropically. Hence we assume, in the limit where length scale effects are unimportant, an isotropic effective material behaviour with 
\begin{align}
\fsigma &=  \tens[4]{C}_{E,\nu} : \feps,\\
\fS &= 2 \Gc  \mepsR,
\end{align}
where $\tens[4]{C}_{E,\nu}$ is the classical isotropic linear elastic stiffness tensor which may be expressed in terms of the Young's modulus $E$ and the Poisson's ratio $\nu$ in standard manner. $\Gc$ is the couple modulus which is a scalar quantity because for the studied 2D structures the $z$-component of the beam vector($l_z=0$) as well as the $x$- and $y$-components of the rotation vector ($\omega_x=\omega_y=0$) are zero -- hence, there is only a single axis of rotation. From Eq. \eqref{eq:couple_stress} it follows that the non-zero components of the couple stress  $\fmu$ have the same indices as the non-zero entries of the rotation gradient~$\grad{\omega}$:
\begin{align}
\fmu &= \mu_{zx} \fe_z \otimes \fe_x + \mu_{zy} \fe_z \otimes \fe_y ,\\
\mgomega &= \omega_{z,x} \fe_z \otimes \fe_x  + \omega_{z,y} \fe_z \otimes \fe_y.
\end{align}
With that a general, anisotropic relation between the couple stress and the rotation gradient is
\begin{align}
\begin{pmatrix}
\mu_{zx}\\
\mu_{zy}
\end{pmatrix} &= 
\underbrace{ \begin{pmatrix}
\gamma_1 & 0\\
0 & \gamma_2
\end{pmatrix}}_{\tens{\Gamma}_{\gamma_1, \gamma_2}}
\begin{pmatrix}
\omega_{z,x} \\
\omega_{z,y} 
\end{pmatrix}
\end{align}
where $\tens{\Gamma}_{\gamma_1, \gamma_2}$ is the anisotropic Cosserat stiffness, which depends on the two additional Cosserat coefficients $\gamma_1, \gamma_2$. 
Altogether the parameters which need to be identified are $\mathcal{P} = \{ E, \nu, \Gc, \gamma_1, \gamma_2 \}$. 

As a method for determining the material constants we apply a fitting approach similar to linear least squares. To obtain effective fitted parameters $\hat{\mathcal{P}}$ we consider the ensemble $\mathcal{E}_c$ of all $N_\mrm{c} \times N_\mrm{CO} \times N_\mrm{S}$ control volumes where $N_\mrm{S}$ is the number of microstructure realizations for a given set of parameters, $N_\mrm{CO}$ the number of cut-outs per realization, $N_\mrm{c}$ the number of control volumes in each cut-out. Over this ensemble we seek to minimize a set of cost functions $\Phi$ with respect to $\mathcal{P}$
\begin{align}
 \hat{\cP} = \argmin_\cP \Phi. 
\end{align} 
Because $\tens[4]{C}_{E,\nu}, \Gc, \tens{\Gamma}_{\gamma_1, \gamma_2}$ are independent of each other we can define and minimize cost function for each set of constants separately. As cost functions the differences between measured stresses and the stresses from the multiplication of the elastic constants with the strains are taken and the sum of their squares is  minimized: 
\begin{align}
 \Phi_{\tens[4]{C}} &= \sum_{\mathcal{E}_\mrm{c}} (\fsigma - \tens[4]{C}_{E,\nu}:\feps):(\fsigma  - \tens[4]{C}_{E,\nu}:\feps),\\
 \Phi_{\Gc} &= \sum_{\mathcal{E}_\mrm{c}} (\fS - 2 \Gc \mepsR):(\fS - 2 \Gc \mepsR), \\
 \Phi_{\Gamma} &= \sum_{\mathcal{E}_\mrm{c}} (\fmu - \tens{\Gamma}_{\gamma_1, \gamma_2}:\mgomega):(\fmu - \tens{\Gamma}_{\gamma_1, \gamma_2} :\mgomega).
 \end{align}  
In order to ensure a positive strain energy density for the 2D (plane strain) case, elastic parameters are restricted to 
\begin{align}
 E > 0 &&  -1>\nu>1 && \Gc>0 &&\gamma_1>0 && \gamma_2>0. 
 \end{align}  
 Numerically this is done via an additional penalty term of the form
 \begin{align}
 \Phi^\mrm{pen} = 1000 \cP^4,
 \end{align}
which we add if the elastic parameters are outside of the admissible range.

There are, of course, alternative methods for determining elastic constants. A straightforward modification of the above approach is to consider, in defining the cost functions, strain differences and compliances instead of stress differences and stiffnesses. It turns out that this does not lead to any appreciable changes in the resulting elastic constants. 

A second alternative method for defining cost functions consists in comparing expressions that directly represent elastic energy density contributions, for instance 
\begin{align}
 \Phi_{\tens[4]{C}} &= \sum_{\mathcal{E}_\mrm{c}} \left( \feps : \tens[4]{C}_{E, \nu}:\feps- \fsigma : \tens[4]{C}^{-1}_{E, \nu}:\fsigma \right )^2,\\
  \Phi_{\Gc} &= \sum_{\mathcal{E}_\mrm{c}} \left (\mepsR : 2 \Gc \mepsR- \fS : \frac{1}{2 \Gc} \fS \right )^2, \\
  \Phi_{\Gamma} &= \sum_{\mathcal{E}_\mrm{c}} \left(\mgomega : \tens{\Gamma}_{\gamma_1, \gamma_2} : \mgomega- \fmu : \tens{\Gamma}^{-1}_{\gamma_1, \gamma_2}: \fmu \right )^2.
 \end{align}
This approach yielded broadly similar results which, however, converged much less reliably for strongly disordered systems with high fluctuations in local stresses and strains. Instead of finding the parameters for the effective, averaged systems it might be possible to determine parameters (stiffnesses or compliances) locally for each control volume and then average over all obtained local elastic constants. A problem of this method is that one cannot use the global statistical symmetries of the material and can, therefore, not determine local elastic constants separately for the different loading cases. We leave a further exploration of such direct averaging methods as an open question for future studies.

\section{Results}
\begin{figure*}
\centering
\includegraphics[width=0.8\textwidth]{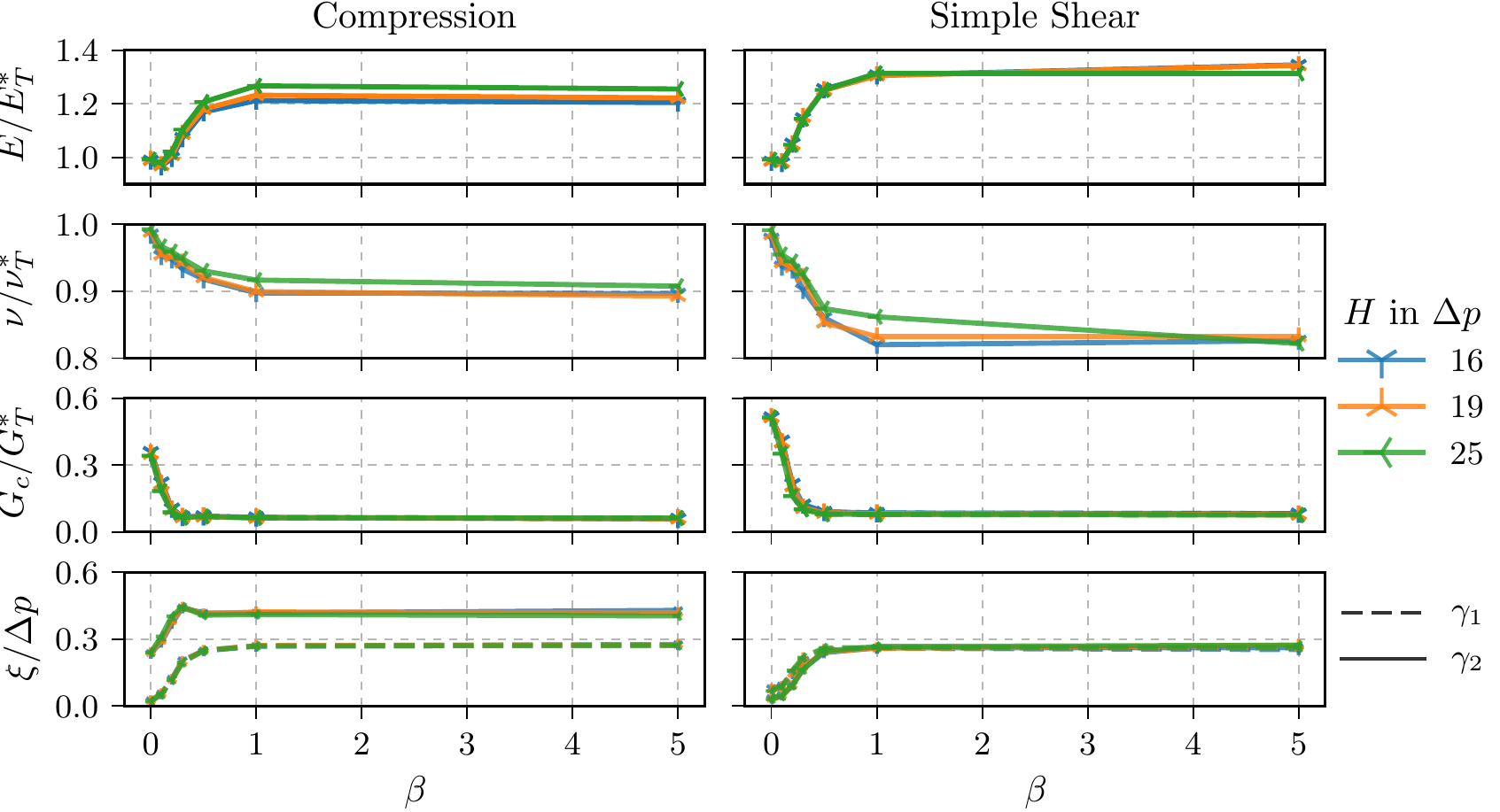}
\caption{Fitted material parameters normalized by the theoretical values for regular honeycombs; internal lengths $\xi_i$ in units of the cell size; data for compression (left) and for simple shear (right). }
\label{fig:material_props}
\end{figure*}

As a benchmark for our results we use the well established analytical solutions for regular, periodic honeycombs of \cite{gibson1999}. For beams which can deform by bending, axial and shear deformation and have the material properties  $\rho_{\mrm{rel}}=0.1, E_{\mrm{B}}=1\cdot 10^8\mrm  \, \mrm{Pa}, $ and $\nu_{\mrm{B}}=0.3$ as used in our simulations, the theoretical Young's modulus, Poisson's ratio and shear modulus  are
\begin{align}
\label{eq:theo_vals}
 E^*_{\mrm T} & \approx 1.44 \cdot 10^5 \, \mrm{Pa}, \nonumber\\
 \nu^*_{\mrm T} & \approx 0.971, & G^*_{\mrm T} \approx 3.65 \cdot 10^4  \, \mrm{Pa}. 
\end{align}

The Cosserat coefficients $\gamma_i$  can be reinterpreted in terms of internal lengths $\xi_i$ \citep{diebels2002}: $\xi_i = \sqrt{\gamma_{i}/2 \Gc} $ with $i=1,2$.  The results for three different sizes are shown in \figref{fig:material_props}. For all sizes the results are very similar which indicates that the computed parameters represent geometry independent material properties. For regular systems, the Young's modulus and Poisson's ratio match very well the analytical solution. As also observed by \citet{zhu2001} and \citet{liebenstein2017}, increasing irregularity increases the elastic moduli of bulk systems, which can be observed for both loading cases. Poisson's ratio decreases with higher randomness, similar to the result reported by \citet{zhu2006}.

\begin{figure*}
\centering
\includegraphics[width=0.8\textwidth]{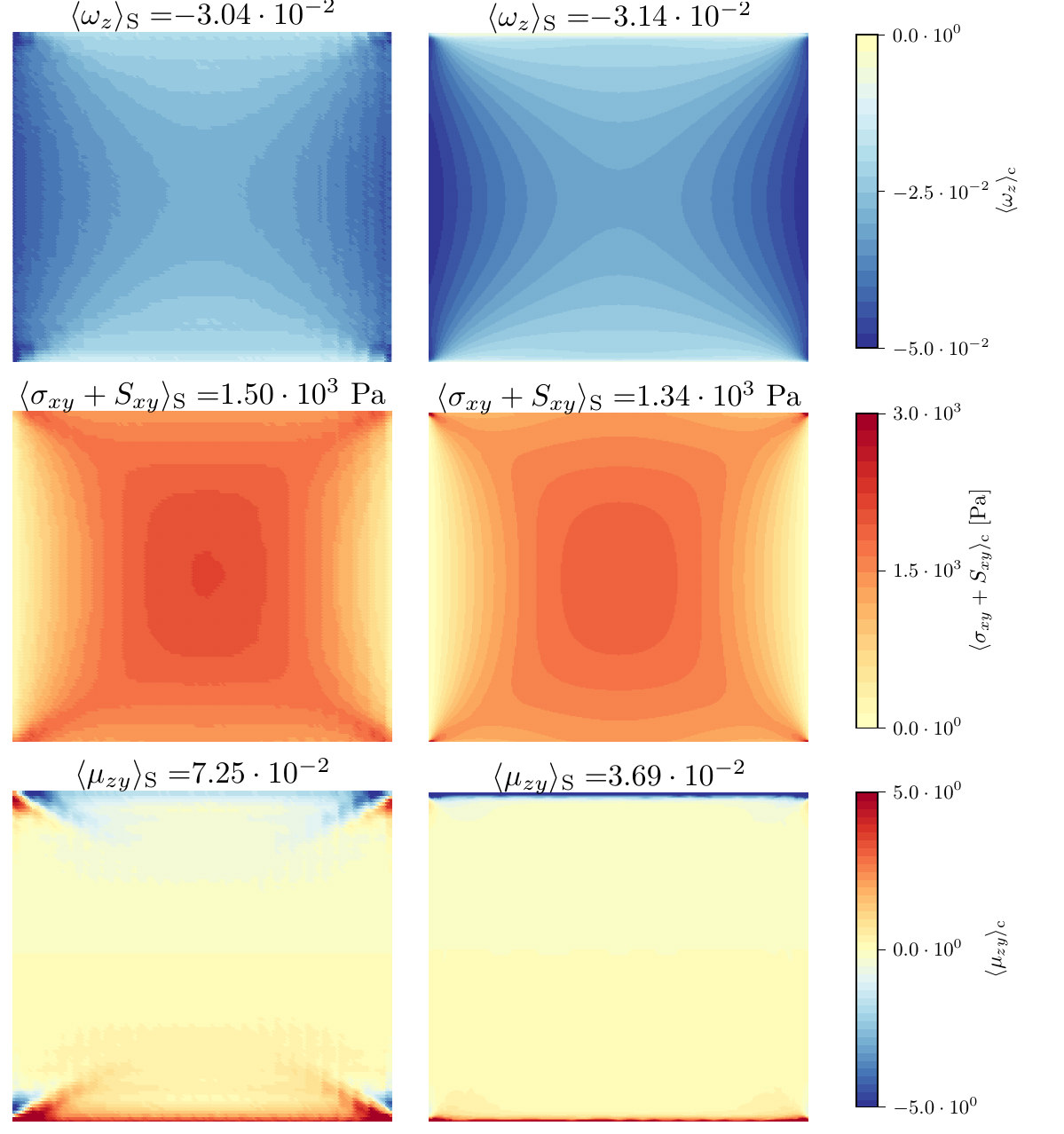}
\caption{Comparison of the rotations, stresses and couple stresses of the averaged beam system (left) and the Cosserat continuum (right) under simple shear, sample parameters $H=19\Delta p$ and $\beta=0$.}
\label{fig:comparison}
\end{figure*}

For both loading conditions the couple modulus $\Gc$ decreases with increasing degree of irregularity, until the irregular systems have a couple modulus of about $\Gc \approx 0.08 G^*_{\mrm T}$. These results are consistent with experimental data quoted in a recent review by \citet{hassanpour2017} who indicate values in the range  of $\Gc \approx 0.1 G^*_{\mrm T}$ for polystyrene foams, syntactic foams and a polymethacrylimide foam. The internal length $\xi$ shows an anisotropic behaviour for regular systems. For both loadings considered, one of the coefficients is almost zero, whereas the other one is of the order of $0.1 \Delta p$. With increasing irregularity both internal length parameters converge to a common value of about $ \xi \approx 0.3 \Delta p$. Higher values of about one cell size ($\Delta p$ in our case), as indicated by the experimental data of \citet{lakes1991, McGregor2014,  hassanpour2017}, are not consistent with the couple stress patterns that derive from our microscale model. We will demonstrate this below. 

For validation we use a direct comparison between averaged results from our micro-scale beam model and a finite element implementation of the Cosserat continuum that uses the material parameters of \figref{fig:material_props}, averaged over all loading modes and system sizes. Macroscopic size effects show only if $\grad \tens{u} \neq \grad \tens{u} \T$, which is not the case for uniaxial compression. (Note that it is nevertheless possible to determine internal length scale parameters by connecting the fluctuating {\em local} variables in the individual realizations - which is what we have done above in case of compression). Therefore, only simple shear loading of the Cosserat medium is considered for validation purposes. We simulate a thin strip with the same width and height as the beam network and a thickness of $D=0.05 H$. By using only 1 element for the thickness, this loading state is similar to a plane stress loading. The intrinsic length requires a mesh which captures the thickness of the boundary layer. Thus the mesh is locally refined towards the surfaces such that the element width is significantly smaller than the internal length. The same boundary conditions as for the beam network are chosen (cf. Table~\ref{tb:Dirichlet BC}) i.e. no in-plane Cosserat rotations and no $x$-displacements at the top and bottom and a $y$-displacement at the top which induces a engineering strain of $\varepsilon^*_{xy} = 5 \%$. The $y$-displacement at the bottom boundary is set to zero whereas the out-of-plane rotations and displacements are unconstrained everywhere else. 

As one can see in \figref{fig:comparison} the stress distribution and the distribution of the rotations match very well. The maximum rotations are found along the free surfaces and gradually decrease towards the inner part of the specimen. The shear stresses show a characteristic distribution similar to the distribution for a classical continuum. From the free surfaces they gradually increase towards the center. The overall average of the continuum system slightly under-predicts the average response.

\begin{figure*}
\centering
\includegraphics[width=0.8\textwidth]{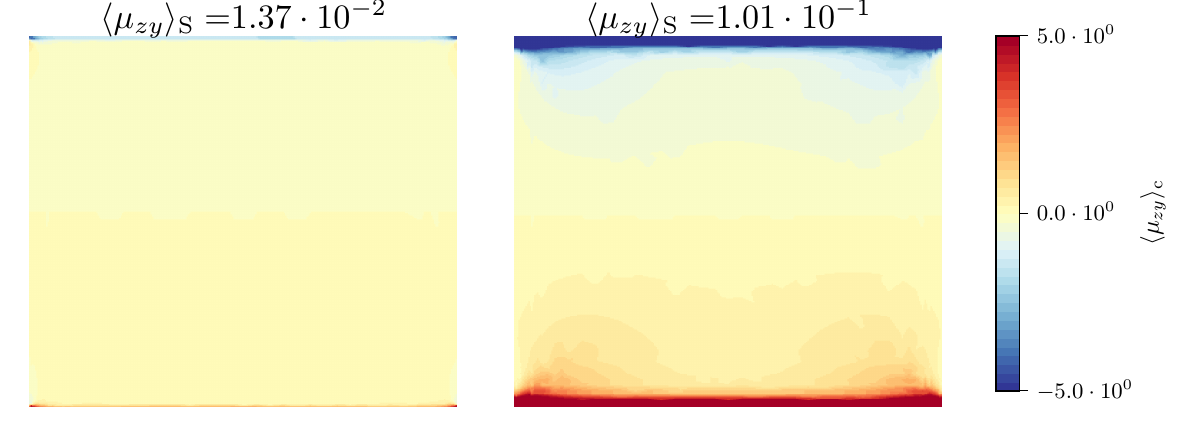}
\caption{Comparison of the couple stresses of the continuum system  with $\bar{l} = 0.5 l$ (left) and $\bar{l} = 2 l$ (right) for $H=19\Delta p$ and $\beta=0$.}
\label{fig:different_length}
\end{figure*}

\begin{figure*}
\centering
\includegraphics[width=0.8\textwidth]{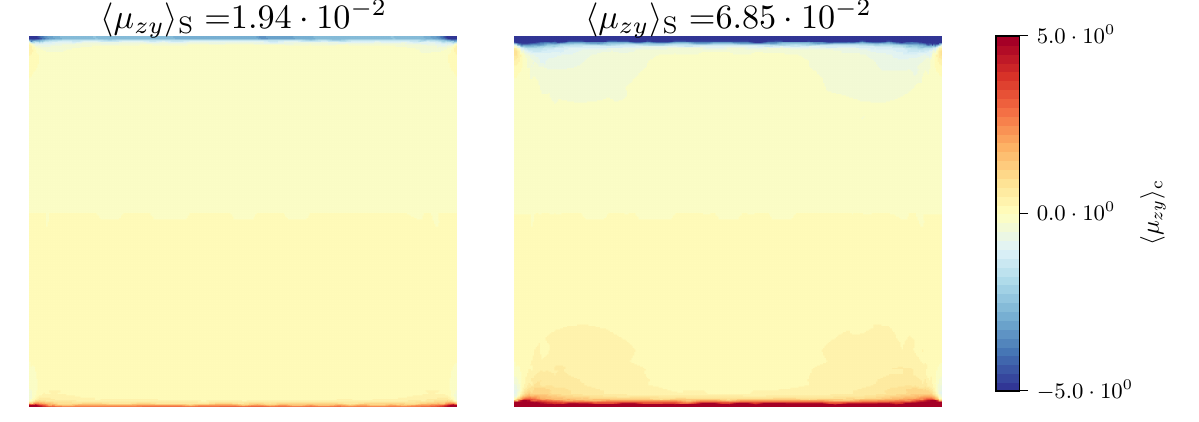}
\caption{Comparison of the couple stresses of the continuum system  with $\bar{G}_\mrm{c} = 0.5 \Gc$ (left) and $\bar{G}_\mrm{c} = 2 \Gc$ (right) for $H=19\Delta p$ and $\beta=0$.}
\label{fig:different_gc}
\end{figure*}

Couple stresses are concentrated in boundary layers at the top and bottom edge of the samples. Both simulations show a comparable thickness of these boundary layers. Because the couple stresses $\fmu$ are linearly dependent on the rotation gradient $\mgomega$, the continuum and beam systems match badly near the specimen corners where both micro-rotations and micro-rotation gradients are high and rapidly alternating. In this region the continuous solutions of the Cosserat equations cannot easily match the continuized fields that by construction are piecewise constant over control volumes of finite size. In general the continuum system shows a bit higher rotations in the boundary layer. Nonetheless the thickness of the layer matches. For comparison \figref{fig:different_length} shows simulations where the internal length was changed by a factor of two in both directions while the other material parameters were kept fixed. As one can see the boundary layer is too narrow for the halved internal length and too wide for the doubled internal length. So even though the width and intensity of the boundary layer do not match perfectly, the high sensitivity of the results to the internal length suggests that the fitted length is close to an optimal solution. Furthermore it is clear from the picture that an internal length of about one cell size would significantly overestimate the influence of the couple stresses. On the other hand, \figref{fig:different_gc} shows, that a couple modulus of $\bar{G}_\mrm{c} \approx 1.5$--$2 \, \Gc$ might provide an even better match. Nonetheless the obtained parameter combination gives a reasonable approximation of the beam network and can be regarded as a informed way to determine Cosserat material properties.

\begin{figure*}
\centering
\includegraphics[width=0.95\textwidth]{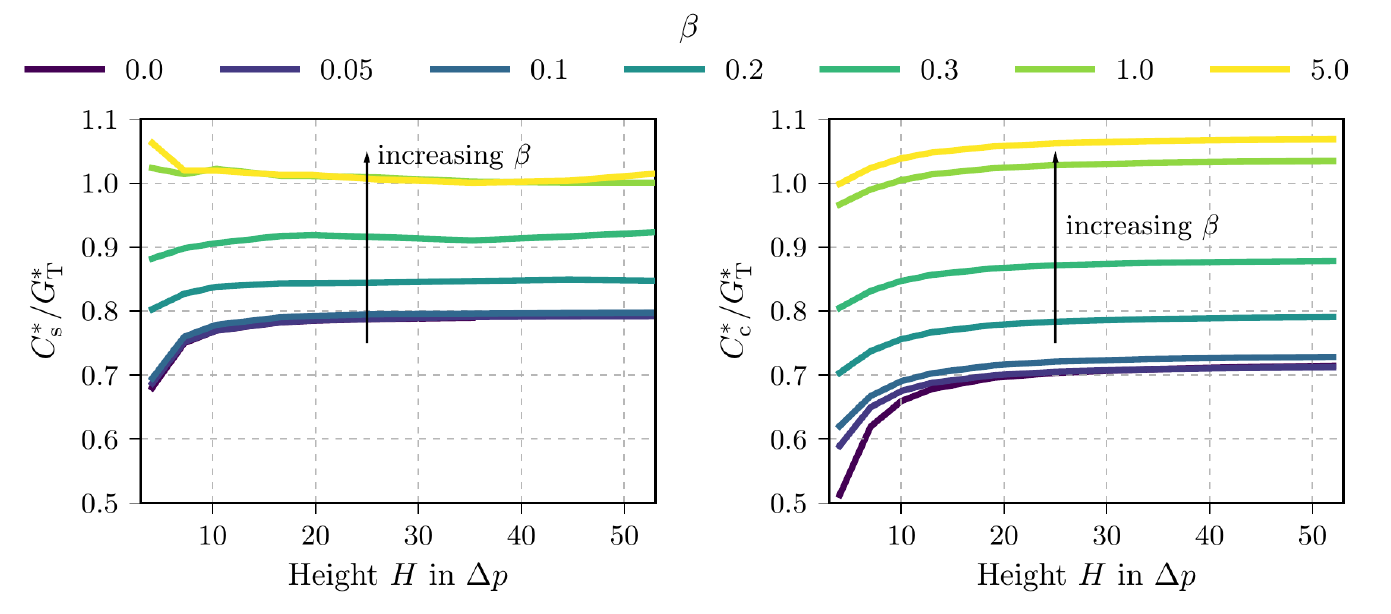}
\caption{Normalized system responses for regular systems ($\beta=0$) of different aspect ratios; left: 
beam models, right: Cosserat continuum models.}\label{fig:system_stiffness}
\end{figure*}

Finally we compare the effective system response of the two systems. 
With the effective engineering strain
\begin{align}
\varepsilon_{xy}^* &= \frac{|u_x(y=H)- u_x(y=0)|}{H}.
\label{eq:eng_strain}
\end{align}
an effective shear stiffness can be defined for the beam network as
\begin{align}
C^*_{\mrm B} &= \frac{1}{A \varepsilon^*_{xy}} \sum_{k=1}^{N_\mrm{R}} R^k_{x} \label{eq:reaction_force_beam}
\end{align}
where $R_x$ are the load induced reaction forces in $x$ direction of all the beams $N_\mrm{R}$ at the loaded surface $A$. In analogy the effective system response of the continuum is
\begin{align}
C^*_{\mrm c} &= \frac{1}{A  \varepsilon^*_{xy}} \int_A t_x \dif A.
\label{eq:reaction_force_conti}
\end{align}
In \figref{fig:system_stiffness} the averaged effective stiffness of all beam systems is for different irregularity parameter $\beta$ compared with result for the corresponding, fitted continuum models. The main trends, namely (i) the increase of overall stiffness with increasing degree of disorder and (ii) the existence of a size effect in the sense that systems of reduced height show a softer response, are well reproduced by the continuum model, though both the size effect and the disorder-induced stiffening are slightly more pronounced in the continuum model. 

\section{Discussion and Conclusions}
We presented a two-scale modelling approach for irregular cellular solids. On the micro-scale, we use beam models - an approach which implies a significant degree of idealization but is computationally much more efficient than for example modelling the microstructure with (Cauchy-) continuum elements. Computational cost is an important factor when dealing with strongly disordered microstructures where load is internally distributed in a strongly heterogeneous manner through force transmission chains \citep{liebenstein2017}: for small sample sizes as considered here, such force transmission chains may span the entire sample and their stochastic character causes significant sample-to-sample variations. Hence, an efficient computational scheme is indispensable in order to capture the statistics of variations and to reliably determine the average behavior of samples with a given degree of irregularity. Despite the large sample-to-sample variability, the computational efficiency of beam models allows us to determine not only macroscopic stiffnesses but also smooth stress, strain, and rotation fields by averaging over sufficiently large ensembles of typically 6000 microstructure realizations for each system size and irregularity.

Our next step is to use the data from the micro-scale beam models in order to parametrize a macroscopic, Cosserat-type continuum model. For such scale transitions typically two approaches are used. In top-down, reverse modelling approaches, one uses the micro-scale model in very much the same manner as one would deal with experimental data: One measures the (here size-dependent) system-scale response and then tries to identify macroscopic parameters that reproduce the same behavior (see e.g. \citet{mora2007, tekoglu2008, diebels2002, diebels_2014}). In bottom-up homogenization approaches one seeks to derive macroscopic, effective materials properties through averaging the micro-heterogeneous material response over a representative volume element (RVE). In the context of the homogenization of a lattice or beam model towards a higher-order continuum on the macro scale this was done for example by \citet{askar1968, bavzant1972, kumar2004, dos2012, elnady2017, trovalusci2017}.
 We note that these methods may be problematic in structurally disordered materials which exhibit strongly heterogeneous stress and strain patterns with long-range correlations. This is the case in our micro-scale simulations: As demonstrated by \citet{liebenstein2017}, stress transmission in the strongly disordered beam models (high disorder parameter $\beta$) occurs through correlated stress transmission chains. In the regime of system sizes where size effects are relevant, in other words in the range of interest for a higher-order continuum model that incorporates internal length scales, these correlated stress transmission chains span the entire simulated sample. Since these chains are of a stochastic nature, this leads to very significant sample-to-sample fluctuations in the macroscopic response. In other words, the simulated micro-sample sizes themselves are below the RVE scale. 

The method we use is different - we use a bottom-up approach that, however, does not implicate a RVE. First we map, in an energetically consistent manner that assumes the Cosserat continuum structure, the stress and deformation fields on the discrete beam level onto spatially continuous fields defined on control volumes that are characteristic of the particular microstructure. Determination of stresses and deformations is independent, and elastic constants are identified by seeking an optimum match between stress and conjugate strain variables through linear mappings. Importantly, this method uses only local information, hence, the RVE problem does not come into play. A remarkable feature of the method is that it allows to determine internal length scale parameters even in deformation settings where we do not expect macroscopic size effects, such as in uniaxial compression. The reason is that the local stress state is, owing to the micro-scale disorder of the material, always multi-axial such that rotation effects are present even in deformation settings that are macroscopically rotation free. Indeed the material constants we determine from the local fields in compression tests are in good agreement with those from simple shear tests. 
When used to evaluate size dependent response observed in shear, these parameters perform well even though they have not been fitted to reproduce shear data. The fact that one can use different uniform or non-uniform deformation settings to probe for the same parameters and then cross-validate them allows to ensure one is indeed dealing with material parameters and not parameters that characterize a specific deformation setting only. 

Even though higher order continuum theories have their roots already 120 years ago they are still not commonly used in engineering applications. One reason is the difficulty to reliably determine constitutive parameters, of which in general theories there may be many. Once increasing computational power allows to efficiently simulate large systems and/or large ensembles of systems that are obviously inaccessible by experiment, this disadvantage may become less serious. In the presented work we showed, in a proof-of-concept study, how Cosserat material parameters can be deduced from simulations of 2D cellular structures. The parameters are obtained by a two step procedure. First the beam network is represented by an equivalent Cosserat continuum. From that the parameters are identified by a simple linear least square optimization approach, relying on local information only. The whole procedure is completely automated, so that no fitting or adapting of parameters by hand is needed. Furthermore it is independent of periodicity or regularity with respect to microstructure and boundary conditions. Of course there remain open questions. 

The present study has focused on statistically isotropic or near-isotropic systems with a single elementary length scale (the cell size). The method we propose may have limitations when applied to anisotropic microstructures where there can be two different sources of anisotropy: Anisotropy on the level of the individual microstructure element (e.g., an anisotropic average cell shape), and anisotropy in the arrangement of different elements, which  might depend on additional correlation lengths (e.g., the shape of regions of higher-than-average or lower-than-average local density). The latter features cannot be captured by an approach which relies on information on the smallest microstrucural length scale, namely the individual cell or control volume, alone and may thus necessitate more sophisticated approaches. Further studies are also needed in order to investigate how the proposed method can be extended to three-dimensional open- or closed-cellular structures.


\begin{backmatter}

\section*{Competing interests}
The authors declare that they have no competing interests.

\section*{Author's contributions}
S.L. devised the continuization scheme, set up and performed all simulations and statistical analysis, and drafted the first version of the manuscript. M.Z. assisted in the final formulation of the continuization scheme and suggested the fitting scheme for determination of elastic parameters. Both authors jointly wrote the final manuscript. 

\section*{Acknowledgements}
The authors acknowledge funding by DFG under Grant no. 1 Za 171-9/1. M.Z. also acknowledges support by the Chinese government under the Program for the Introduction of Renowned Overseas Professors (MS2016XNJT044).


\bibliographystyle{bmc-mathphys} 
\bibliography{bibfile}     
\nocite{label}


\end{backmatter}
\end{document}